\documentclass[TinyML]{acmart}

\AtBeginDocument{%
  \providecommand\BibTeX{{%
    \normalfont B\kern-0.5em{\scshape i\kern-0.25em b}\kern-0.8em\TeX}}}

\setcopyright{rightsretained}
\copyrightyear{2024}
\acmYear{2024}

\begin{document}

\title{SpokeN-100: A Cross-Lingual Benchmarking Dataset for The Classification of Spoken Numbers in Different Languages}


\orcid{0000-0002-3405-1311}
\author{René Groh}
\email{rene.groh@fau.de}
\affiliation{%
  \institution{Department Artificial Intelligence in Biomedical Engineering}
  \streetaddress{P.O. Box 1212}
  \city{Erlangen}
  \state{Bavaria}
  \country{Germany}
  \postcode{43017-6221}
}

\author{Nina Goes}
\affiliation{%
  \institution{Department Artificial Intelligence in Biomedical Engineering}
  \streetaddress{1 Th{\o}rv{\"a}ld Circle}
  \city{Erlangen}
    \state{Bavaria}
  \country{Germany}}

\author{Andreas M Kist}
\email{andreas.kist@fau.de}
\affiliation{%
  \institution{Department Artificial Intelligence in Biomedical Engineering}
  \city{Erlangen}
  \state{Bavaria}
  \country{Germany}
}

\renewcommand{\shortauthors}{Groh et al.}

\begin{abstract}
    Benchmarking plays a pivotal role in assessing and enhancing the performance of compact deep learning models designed for execution on resource-constrained devices, such as microcontrollers. Our study introduces a novel, entirely artificially generated benchmarking dataset tailored for speech recognition, representing a core challenge in the field of tiny deep learning. SpokeN-100 consists of spoken numbers from 0 to 99 spoken by 32 different speakers in four different languages, namely English, Mandarin, German and French, resulting in 12,800 audio samples. We determine auditory features and use UMAP (Uniform Manifold Approximation and Projection for Dimension Reduction) as a dimensionality reduction method to show the diversity and richness of the dataset. To highlight the use case of the dataset, we introduce two benchmark tasks: given an audio sample, classify (i) the used language and/or (ii) the spoken number. We optimized state-of-the-art deep neural networks and performed an evolutionary neural architecture search to find tiny architectures optimized for the 32-bit ARM Cortex-M4 nRF52840 microcontroller. Our results represent the first benchmark data achieved for SpokeN-100.
\end{abstract}


\maketitle

\section{Introduction}
In recent years, deep neural networks have experienced a significant surge, with numerous studies focusing on scaling up model sizes and data volumes to develop robust artificial intelligence (AI) models for broad applications \cite{min2023recent, yang2023diffusion}. While these models, which require significant computational resources and data, dominate the AI landscape, specific tasks for edge computing, such as the field of tiny deep learning (tinyDL), have received comparatively less attention. TinyDL aims to harness the power of deep neural networks within the severe limitations of microcontroller hardware and represents a unique intersection of advanced neural network techniques and ultra-low-power computing \cite{reddi2021widening, zaidi2022unlocking, capogrosso2023machine}. 

To address the challenges posed by these highly constrained hardware environments, significant strides have been made across multiple fronts. These include the development of dedicated hardware, the creation of innovative software solutions such as AI inference frameworks, and the optimization of neural architectures for enhanced performance on limited resources. Emerging from these efforts is a dynamic field of research known as Neural Architecture Search (NAS), which focuses on automating the design of neural network models \cite{liberis2021munas, banbury2021micronets, lin2020mcunet}. 

Comprehensive benchmarking datasets are crucial for effectively evaluating the performance of individual deep learning models and emerging NAS algorithms. The MLperf Tiny Benchmark \cite{banbury2021mlperf} offers a specialized suite of benchmark datasets designed specifically for the requirements of TinyDL, providing a robust framework for evaluating both NAS algorithms and deep learning models in this domain. Included in the MLperf Tiny Benchmark is the Speech Commands dataset \cite{warden2018speech}, a speech recognition dataset that encompasses spoken numbers from 0 to 9, previously analyzed in \cite{ouisaadane2019english}. Additionally, the AudioMNIST dataset \cite{becker2023audiomnist} features numbers from 0 to 9 in the English language. There are several more datasets featuring spoken numbers in different languages, e.g. Arabic \cite{wazir2019spoken} or Bangla \cite{sen2022novel}. However, to our knowledge, there exists no dataset covering spoken numbers from 0 to 99 in English, German, French, and Mandarin, spoken from the same set of speakers for different languages. 

SpokeN-100 aims to fill the aforementioned gap and enrich the benchmarking dataset landscape, with a specific focus on advancements in the TinyDL domain. This dataset is unique due to its heterogeneity, encompassing spoken numerical utterances from a variety of speakers across four languages: English, German, French, and Mandarin. The data in this study is entirely artificially generated using advanced AI models, ensuring a controlled yet realistic variety in the dataset. This is in contrast to crowd-sourced data, which can have an uncontrollable bias due to factors such as background noise. The analysis shows that SpokeN-100 is highly diverse, making it an ideal resource for a variety of learning tasks within TinyDL. This dataset is valuable for evaluating and enhancing speech recognition algorithms and serves as a testbed for exploring the capabilities of TinyDL models in handling multilingual and diverse acoustic environments.

\begin{figure*}[h]
   \centering
   \includegraphics[width=\textwidth]{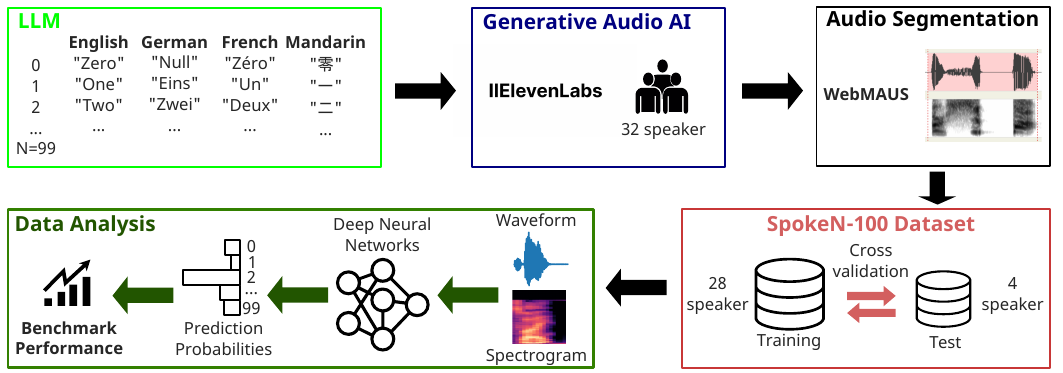}
   \caption{Overview of this study as a flow diagram.}
   \label{fig:overview}
\end{figure*}

\section{Methods}
In this chapter, we detail the development of the SpokeN-100 dataset (refer to Fig. \ref{fig:overview}), elaborate on the measures implemented to guarantee data quality and describe the methodologies employed for establishing initial benchmark results.

\subsection{Dataset Creation}
In the initial phase, we employed a Large Language Model (LLM), specifically ChatGPT (GPT 3.5), to generate textual representations of spoken numbers across four languages, for example 42 would be forty-two in English and zweiundvierzig in German. Subsequently, we utilized ElevenLabs \cite{elevenlabs2023}, a text-to-speech service using generative AI models, for converting these texts into spoken words. From ElevenLabs' array of predefined speakers, we carefully selected 32 speakers—comprising an equal split of 16 female and 16 male voices—to ensure a gender-balanced representation in the audio dataset. This process involved creating texts for numbers ranging from 0 to 99 in each language and generating their corresponding audio renditions using the ElevenLabs API, with each audio sample recorded at a $44.1 kHz$ sampling rate. Predominantly, we relied on ElevenLabs’ state-of-the-art multilingual model v2. However, due to the subpar audio quality of some speakers, we occasionally resorted to the earlier multilingual model v1. \\
To segment our audio into the designated 100 classes, we employed the WebMAUS web application, which synchronizes speech recordings with their corresponding text transcriptions \cite{kisler2012signal}. This approach was used for the Indo-European languages (English, German, and French) to accurately determine the timestamps for the start and end of each number. In French, however, certain compound numbers like '21' (vingt et un) presented a unique challenge, as WebMAUS typically identifies timestamps for individual words. To address this, we manually grouped such numbers, selecting the beginning and end of the entire sequence instead. Utilizing these timestamps, we were able to divide the audio data into smaller segments, each representing one of the 100 classes. To prevent abrupt endings, we incorporated a 0.2-second pause at the end of each audio clip.\\
For the Mandarin dataset, where WebMAUS was unsuitable, we adopted a different strategy. We analyzed the rolling standard deviation of the audio signal to pinpoint the start and end of the numbers. After normalizing this rolling standard deviation, we set a threshold value at $0.05$ to effectively delineate the numbers. 
We or native speakers of the four languages listened to the individual audio samples to ensure their integrity. 

\subsection{Data quality and diversity between languages and speakers}
To ensure our benchmarking dataset was diverse and comprehensive, we undertook detailed data analysis. We assessed the data quality using descriptive statistics and manifold learning. For the latter, we relied on UMAP for a dimensionality reduction technique to enhance the visual inspectability of the dataset.

\subsubsection{Calculation of the fundamental frequency}
An important feature for describing human speech is its pitch, which is closely related to the fundamental frequency $F_0$ \cite{pisanski2021human}. The pitch describes the perceived frequency of a sound and is a key feature in identifying and distinguishing voices. It is crucial for analyzing speech patterns and plays an important role in linguistic prosody, which influences the meaning and emotions conveyed in spoken language. \\
To verify and uphold the quality of our artificially generated audio data, we employed the PYIN algorithm \cite{mauch2014pyin} as part of the librosa Python library \cite{mcfee2015librosa} to compute $F_0$ for each of the 12,800 audio samples. We downsampled the audio data to $11025 Hz$, because lower sampling rates enhance the reliability of the algorithm. For our analysis, we chose a frame length of 512 data points, alongside a singular threshold setting for peak estimation. To ensure accuracy in our $F_0$ measurements, we excluded all frequencies outside the typical vocal range, specifically those below $50 Hz$ and above $300 Hz$, as the fundamental frequency predominantly resides within this frequency band \cite{titze1998principles}. In the final step, we calculated the mean fundamental frequency ($\bar{F_0}$) for each audio file by averaging the remaining frequencies.

\subsubsection{Manifold learning using UMAP}
UMAP is a widely used technique for manifold learning and dimensionality reduction. It is known for its efficiency in dealing with large data sets and its ability to preserve local structure in high-dimensional data \cite{mcinnes2018umap}. We used UMAP to gain more insight into the diversity of the SpokeN-100 dataset. Firstly, we downsampled each audio file to 8000 data points using the Fourier method, effectively reducing the temporal resolution while retaining essential spectral information. Subsequently, we computed the mel-frequency spectrogram ($n\_fft=1024$, $hop\_length=256$, $n\_mels=128$), a well-established time-frequency representation of the audio signal. The mel-frequency spectrogram offers a frequency axis scaled to align with human auditory perception, enhancing our ability to capture relevant acoustic features and patterns in the data. We converted the power mel-spectrogram to decibel units (dB) and used the resulting features to obtain a two-dimensional embedding using UMAP.

\subsection{Classification using deep neural networks}
We established two distinct classification tasks using the full SpokeN-100 dataset. Given an audio sample file, (i) identify the language that is spoken or (ii) identify which number (0 ... 99) is articulated. That means it is a four- and a one-hundred-class classification problem, respectively. Random guessing would therefore yield a 25 \% and 1 \% accuracy, respectively. Outperforming these values means that a given model is guessing better than a random guess and indicates successful learning. We employed the same pre-processing methods that were utilized for the UMAP embedding. For our baseline neural architectures, we selected a diverse range of architectural designs. Initially, we crafted three straightforward networks: a one-dimensional convolutional neural network (CNN), a two-dimensional CNN, and a recurrent neural network (RNN) based on Bidirectional-LSTM layers. Additionally, we trained an EfficientNet-B0 \cite{tan2019efficientnet} and a multi-head transformer network \cite{vaswani2017attention}. All these networks used mel-spectrogram features as input, with the exception of the one-dimensional CNN, which processed normalized waveform inputs. All models were trained within the TensorFlow framework v. 2.13 with the Adam optimizer, each using an empirically tuned learning rate scheduler. More details about the architectures and the learning rate schedulers can be found in Appendix \ref{appendix:baseline_models}. Training was performed on a workstation equipped with an A5000 GPU with 24 GB of memory for 75 epochs for each network.

\subsection{Neural architecture search and deployment on microcontroller}
To identify neural architectures suitable for microcontroller deployment, we conducted an evolutionary neural architecture search using the EvoNAS framework, as detailed in \cite{groh2023end}. Briefly, EvoNAS uniquely outputs TensorFlow Lite (TFLite) Micro neural network architectures that are directly deployable. The search space was thus designed to accommodate this. Each network processes an 8000 data point waveform input, followed by a short-term Fourier transform (STFT) using the kapre library \cite{choi2017kapre} that wraps the STFT implementation into a TensorFlow layer as the TensorFlow version lacks certain necessary operations when deployed using TFLite Micro. The EvoNAS optimization algorithm was executed for 15 generations. Each generation contained a population with 100 competing and evaluated individuals. To find a solution to the multi-objective optimization problem, we calculated the fitness $F$ for each individual neural network architecture by using the following equation: \\

\begin{math}
  F = \alpha \cdot A + \beta \cdot \frac{R_{min}}{R} + \gamma \cdot \frac{E_{min}}{E},
\end{math} \\

where $A$ is the accuracy of the validation dataset, $R$ the obtained rom usage (flash memory) in Bytes (B), and $E$ is the energy consumption in Millijoule (mJ). The latter already contains the inference information as it is calculated based on it. However, the inference time and power consumption during inference are measured directly on-device by deploying each neural network to the microcontroller. To normalize the measurement into the range from 0 to 1, we determined the minimum possible value for the rom usage and energy consumption ($R_{min}=170000 B$, $E_{min}=1.0$). In addition, the scaling parameters $\alpha = 0.6$, $\beta=0.2$ and $\gamma=0.2$ were set so that the algorithm favors certain objective measures more than others.

\subsection{SpokeN-100 as a TinyDL benchmark dataset}
SpokeN-100 features a diverse array of 32 speakers identified by their ElevenLabs name. For validation reasons, we performed multiple times stratified dataset splits across speakers. SpokeN-100 contains a plain text file detailing the data splits according to speaker identity (see Appedix \ref{appendix:code}). These splits encompass eight distinct groups, each consisting of four speakers. For evaluating the final performance on the dataset, models are trained using 28 speakers, while the remaining four speakers in each split are utilized to assess the test accuracy (see Fig. \ref{fig:overview}). The measure of performance of one model is determined by calculating the average test accuracy across all eight data splits.

\section{Results}
In this section, our data analysis demonstrates the richness and diversity of the SpokeN-100 dataset, highlighting its complexity despite being entirely synthetically generated. We utilized UMAP for dimensionality reduction, enabling us to effectively visualize the dataset in a two-dimensional space. Following this, we embarked on training multiple deep neural networks and conducting an evolutionary neural architecture search, to establish initial benchmark results for the dataset.

\subsection{SpokeN-100 is a highly diverse AI-Generated Dataset}
\begin{figure}[ht]
   \centering
   \includegraphics[width=\linewidth]{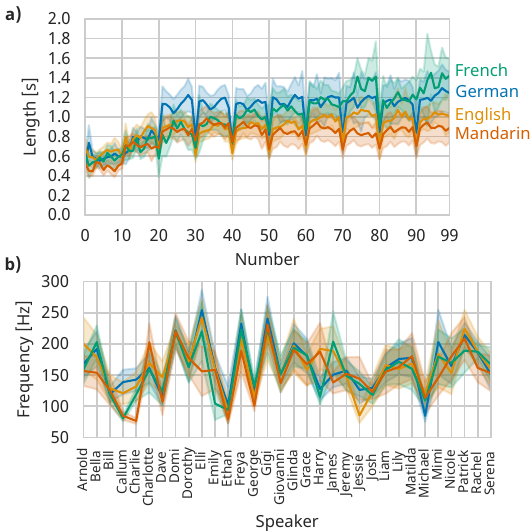}
   \caption{Descriptive statistics of the dataset. a) Mean audio sample length of each spoken number across all speakers for four different languages. The shaded marks the standard deviation. b) Determined fundamental frequencies $F_0$ for each speaker. $F_0$ was calculated for each audio sample and then averaged across all speakers. The shaded area indicates the standard deviation.}
   \label{fig:descriptive_statistics}
\end{figure}

SpokeN-100 is artificially generated which could lead to faulty data and thus influence the performance of machine learning models. We examined the audio lengths for each spoken number sample. Our analysis revealed that in every language, single-digit numbers are generally shorter than double-digit ones (see Fig. \ref{fig:descriptive_statistics}a). We also observed that numbers at tens positions (like 20, 30, etc.) are shorter than other double-digit numbers. This pattern is consistent across languages, as double-digit numbers typically combine a tens word with a single-digit word, such as 'twenty-eight' being a combination of 'twenty' and 'eight'. This is in line with their real counterpart and suggests the validity of SpokeN-100.
To verify consistency among speakers across audio samples and languages in our dataset, we analyzed the fundamental frequency $F_0$ for each audio sample, calculating the average for every speaker. Our results show that $\bar{F_0}$ is almost consistent for each speaker across files and languages (see Fig. \ref{fig:descriptive_statistics}b). Comparing all speakers with each other, it is noticeable that the dataset features different $F_0$ frequencies, indicating that the data is highly diverse. Additionally, the obtained average fundamental frequency values follow the natural range of human language, which is between 50 and 300 Hz. Both findings affirm the dataset's high data quality.

\begin{figure}[ht]
   \centering
   \includegraphics[width=\linewidth]{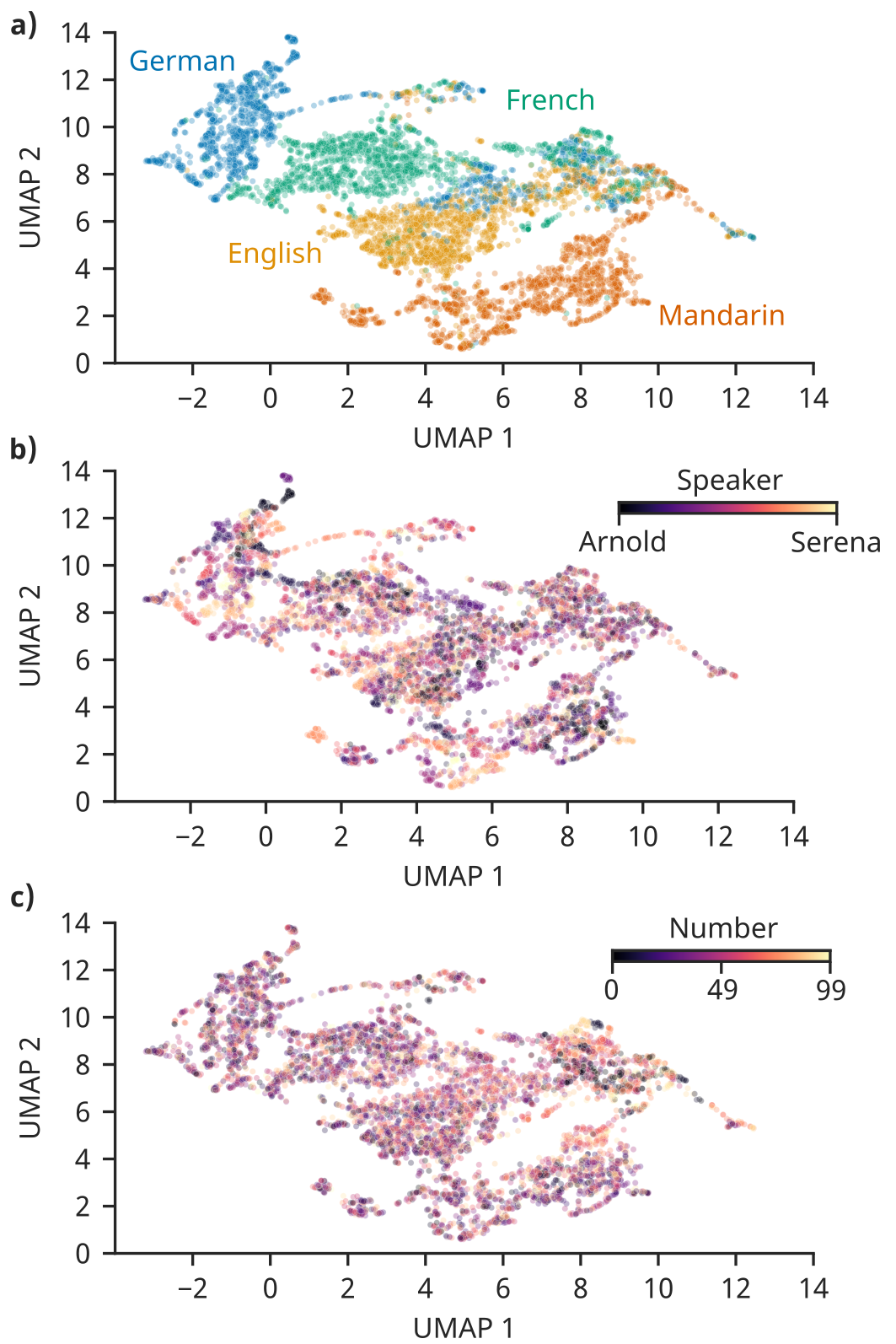}
   \caption{Low dimensional visualization of all audio samples with UMAP. a) UMAP embedding color-coded by language. b) UMAP embedding color-coded by speaker. Each speaker is assigned a value between 0 and 1 to create a gradient visualization. c) UMAP embedding color-coded by spoken number.}
   \label{fig:umap_embedding}
\end{figure}

In our subsequent data analysis, we utilized UMAP for dimensionality reduction. The UMAP results, depicted in Fig. \ref{fig:umap_embedding}a, reveal distinct clustering of data points according to languages. However, when these points are color-coded based on speakers (as shown in Fig. \ref{fig:umap_embedding}b), no clear clusters emerge, suggesting a high level of diversity in speaker selection. Further color-coding of points in the UMAP embedding based on the numbers in each audio sample (Fig. \ref{fig:umap_embedding}c) also fails to show distinct clusters, implying that the task of classifying numbers could be inherently complex.

\begin{table*}
  \caption{Results for language and numbers classification for different deep neural network architectures. All state-of-the-art architectures were non-deployable (n.d.) to the nRF52840 microcontroller. The neural networks in the bottom three rows were successfully optimized utilizing the EvoNAS end-to-end pipeline \cite{groh2023end}. For each network, we provide the mean test accuracy along with its standard deviation, following the previously described data split methodology.}
  
  \label{tab:dnn_results}
  \begin{tabular}{ccccl}
    \toprule
    Architecture & Number of Parameters & Inference Time (ms) & \multicolumn{2}{c}{Test Accuracy}\\
    \cline{4-5}
    & & & Languages & Numbers \\
    \midrule
    2D CNN & 1,891,972 & n.d. & $0.927 \pm 0.017$ & $0.565 \pm 0.054$ \\
    1D CNN & 16,368,340 & n.d.& $0.815 \pm 0.029$ & $0.132 \pm 0.015$ \\
    RNN & 5,709,700 & n.d. & $0.920 \pm 0.015$ & $0.614 \pm 0.041$ \\
    EfficientNet-B0 CNN & 5,365,415 & n.d. & $0.963 \pm 0.011$ & $0.848 \pm 0.037$\\
    Transformer & 1,352,324 & n.d. & $0.945 \pm 0.012$ & $0.729 \pm 0.046$\\
    \bottomrule
    EvoNAS (most accurate) & 18,264 & 1761 & $0.837 \pm 0.022$ & $0.175 \pm 0.025$ \\
    EvoNAS (fastest) & 6,496 & 152 & $0.364 \pm 0.027$ & $0.022 \pm 0.003$ \\
    \textbf{EvoNAS (fittest)} & \textbf{5,379} & \textbf{368} & $\mathbf{0.813 \pm 0.021}$ & $\mathbf{0.126 \pm 0.020}$ \\
  \end{tabular}
\end{table*}

\subsection{SpokeN-100 is suitable as a TinyDL benchmarking dataset}
Evaluating various state-of-the-art neural network architectures using the SpokeN-100 benchmark split, our study (refer to Table \ref{tab:dnn_results}) demonstrates that all networks successfully learned the necessary patterns for the classification tasks. In language classification, models processing mel-spectral features uniformly achieved a mean test accuracy above $90\%$. In contrast, the one-dimensional CNN, which used waveform inputs, underperformed with a mean test accuracy of $81.5\%$. When tackling the more complex task of number classification, involving a one-hundred-class problem, EfficientNetB0 emerged as the top performer. The one-dimensional CNN consistently showed the lowest performance. These findings point to the conclusion that implementing preprocessing, specifically through mel-spectrogram calculation, significantly enhances classification accuracy.

Deploying and executing all the previously mentioned models on a microcontroller is impractical due to their extensive number of parameters and resultant large network sizes. To address this, we conducted an evolutionary neural architecture search (EvoNAS). Table \ref{tab:dnn_results} showcases three models that excelled in terms of accuracy, inference time, and overall fitness (more architectural details in Appendix \ref{appendix:evonas_models}). Each model was successfully deployed on the nRF52840 microcontroller. The most accurate model had a slower inference time of $1.7$ seconds per inference. In contrast, the fastest model experienced a significant decrease in test accuracy. The most balanced, or 'fittest', architecture achieved an inference time of 368 milliseconds, with a test accuracy of $81.3\%$ for languages and $12.6\%$ for numbers, setting the benchmark for SpokeN-100 on a microcontroller.

\section{Discussion}

Our goal was to enhance speech recognition for spoken numbers by introducing a dataset comprising spoken numbers from 0 to 99. Our study demonstrates that the two classification tasks we presented can be effectively addressed using state-of-the-art neural networks. However, these networks become less feasible on highly constrained hardware, leading to compromises that result in diminished performance compared to the previously discussed models. 

To facilitate inference on the microcontroller, we downscaled the input data to 8000 data points, a measure that might induce some information loss. Using larger input sizes would exceed the microcontroller's peak RAM capacity. Additionally, we had to limit the search space, meaning that we could not use the same STFT parameters that we used as pre-processing for the other neural networks because the resulting models would not be too large to deploy. Looking ahead, opting for microcontrollers with slightly more computational capability could be advisable. However, the aforementioned limitations lead to a huge drop in test accuracy when looking at the number classification. Enhancing the classification performance on edge microcontrollers, especially in terms of accuracy, remains an unresolved challenge.

Besides these limitations, it is noteworthy that the architectures discovered by the EvoNAS optimization algorithm are deployable to the microcontroller. These architectures are designed to function 'out of the box' within our search space, eliminating the need for additional data pre-processing on the microcontroller prior to running inference.

We want to highlight the practical relevance that SpokeN-100 can have. A potential use case is in device control scenarios requiring numerical input, such as in robot navigation \cite{huang2023visual} or the operation of wearable devices \cite{wilde2015prototyping}. Looking ahead, we plan to enhance SpokeN-100 by releasing a second version featuring a greater variety of speakers and additional languages, further diversifying the dataset.

\section{Conclusion}
In this study, we presented SpokeN-100 which is a speech recognition dataset of spoken numbers. SpokeN-100 is especially suited to be used as a benchmark for TinyDL algorithms, such as neural architecture search. Other than that we believe that SpokeN-100 also has a high practical relevance. SpokeN-100 provides a starting point to train models to be able to do this. As it encompasses multiple languages, resulting machine learning models are also able to mitigate language barriers. We believe that SpokeN-100 can have a very positive impact on the development of TinyDL applications, further accelerating the field.

\begin{acks}
We receive funding from the Bavarian State Ministry of Science and the Arts (StMWK) and Fonds de recherche du Québec (FRQ) under the Collaborative Bilateral Research Program Bavaria – Québec managed by WKS at Bavarian Research Alliance (BayFOR) and Fonds de recherche du Québec – Santé (FRQS). The presented content is solely the responsibility of the authors and does not necessarily represent the official views of the above funding agencies.
\end{acks}

\clearpage

\bibliographystyle{ACM-Reference-Format}

\appendix

\section{Neural Networks}

\subsection{Baseline models}
\label{appendix:baseline_models}
In the following we provide more details about the used neural architectures. Layers are named after the type of layer, followed by the kernel size, the amount of units and the used activation function, e.g. Conv3-64-relu stands for a convolutional layer with kernel size of 3, 64 units and ReLU as activation function. The final fully connected (FC) layer consists always of four or one-hundred units depending on the classification task (languages or numbers).\\
We trained all networks for 75 epochs using individual learning rate schedulers.

\subsubsection{2D CNN}
Our 2D CNN consists of the following layers: Input(128, 32), Conv3-16-relu, MaxPooling2, Conv3-32-relu, MaxPooling2, Conv3-64-relu, Flatten, FC-256-relu, FC-128-relu, FC-4-softmax. Convolutional layers in this model have two-dimensional kernels.\\
The used learning rate was $0.0001$ until epoch 25, $0.00005$ until epoch 50, and $0.00001$ afterward.

\subsubsection{1D CNN}
Our 1D CNN consists of the following layers: Input(8000, 1), Conv3-8-relu, MaxPooling2, Conv3-16-relu, MaxPooling2, Conv3-32-relu, MaxPooling2, Conv3-64-relu, Flatten, FC-256-relu, FC-128-relu, FC-64-relu, FC-4-softmax. Convolutional layers in this model have one-dimensional kernels.\\
The used learning rate was $0.0001$ until epoch 25, $0.00005$ until epoch 50, and $0.00001$ afterward.

\subsubsection{RNN}
Our RNN consists of the following layers: Input(128, 32), Bidirectional(LSTM-128), Bidirectional(LSTM-256), Bidirectional(LSTM-512), Dropout-0.5, FC-256-relu, FC-128-relu, FC-4-softmax. In this architecture, the first two LSTM layers are returning sequences. \\
The used learning rate was $0.0001$ until epoch 25, $0.00005$ until epoch 50, and $0.00001$ afterward.

\subsubsection{EfficientNetB0 CNN}
The architecture for EfficientNetB0 is described in \cite{tan2019efficientnet}. We used the implementation from TensorFlow (pre-trained on ImageNet) and added as top of the network the following layers: GlobalAveragePooling, FC-1024-relu, FC-4-softmax. \\
The used learning rate was $0.001$ until epoch 25, $0.0005$ until epoch 50, and $0.0001$ afterward.

\subsubsection{Transformer}
Our Transformer network consists of the following layers: Input(128, 32), FC-128-linear, PositionalEncoding, Dropout(0.1), four EncoderLayers, FC-4-softmax. Each EncoderLayer consists of a MultiHeadAttention block, Dropout(0.1), LayerNormalization, FC-1024-relu, FC-128-None, Dropout(0.1) and LayerNormalization. More details can be found in \cite{vaswani2017attention}.\\
The used learning rate was $0.0001$ until epoch 30, $0.00001$ until epoch 50, $0.000001$ until epoch 70, and $0.0000001$ afterward.

\subsection{EvoNAS model}
\label{appendix:evonas_models}
\subsubsection{Fittest Model}
The architecture of our fittest model consists of the following layers: Input(8000, 1), STFT(n\_fft=112, hop\_length=256), Magnitude, Mel-Filterbank (n\_mels=44), DepthwiseConv5, Conv3-8, MaxPooling3, DepthwiseConv3, Conv1-14, Conv5-12, MaxPooling4, Conv2-7, BatchNormalization, GlobalAveragePooling, FC-40-relu, Dropout(0.2), FC-4-softmax.

\subsubsection{Most Accurate Model}
The architecture of our most accurate model consists of the following layers: Input(8000, 1), STFT(n\_fft=96, hop\_length=144), Magnitude, Mel-Filterbank (n\_mels=52), Conv1-6, ReLU, Conv4-27, BatchNormalization, DepthwiseConv2, ReLU, DepthwiseConv3, MaxPooling4, DepthwiseConv2, DepthwiseConv5, Conv4-17, ReLU, Conv3-21, MaxPooling2, BatchNormalization, MaxPooling2, GlobalAveragePooling, FC-48-relu, Dropout(0.1), FC-48-relu, FC-4-softmax.

\subsubsection{Fastest Model}
The architecture of our fastest model consists of the following layers: Input(8000, 1), STFT(n\_fft=80, hop\_length=304), Magnitude, Mel-Filterbank (n\_mels=68), DepthwiseConv2, DepthwiseConv5, BatchNormalization, GlobalAveragePooling, Dropout(0.2), FC-48-relu, FC-120-relu, FC-4-softmax.

\section{Code and dataset}
\label{appendix:code}
The code for data exploration and neural network training can be found at the following link: \url{https://github.com/ankilab/SpokeN-100}. The dataset can be downloaded via \url{https://zenodo.org/records/10810044}.

\end{document}